# End-to-End Mineral Exploration with Artificial Intelligence and Ambient Noise Tomography

Jack Muir[1,2], Gerrit Olivier[1,3], Anthony Reid[1,4]

1. Fleet Space Technologies, Beverley, South Australia, Australia
2. Research School of Earth Sciences, The Australian National University, Canberra, ACT, Australia
3. Centre for Ore Deposits and Earth Sciences, University of Tasmania, TAS, Australia
4. Department of Earth Sciences, The University of Adelaide, Adelaide, SA, Australia

## Abstract

This paper presents an innovative end-to-end workflow for mineral exploration, integrating ambient noise tomography (ANT) and artificial intelligence (AI) to enhance the discovery and delineation of mineral resources essential for the global transition to a low carbon economy. We focus on copper as a critical element, required in significant quantities for renewable energy solutions. We show the benefits of utilising ANT, characterised by its speed, scalability, depth penetration, resolution, and low environmental impact, alongside artificial intelligence (AI) techniques to refine a continent-scale prospectivity model at the deposit scale by fine-tuning our model on local high-resolution data. We show the promise of the method by first presenting a new data-driven AI prospectivity model for copper within Australia, which serves as our foundation model for further fine-tuning. We then focus on the Hillside IOCG deposit on the prospective Yorke Peninsula. We show that with relatively few local training samples (orebody intercepts), we can fine tune the foundation model to provide a good estimate of the Hillside orebody outline. Our methodology demonstrates how AI can augment geophysical data interpretation, providing a novel approach to mineral exploration with improved decision-making capabilities for targeting mineralization, thereby addressing the urgent need for increased mineral resource discovery.

## Introduction

The scale of the global challenge to transition to a low carbon economy is difficult to understate. New technologies are required to accelerate all aspects of the transition, including in the discovery and delineation of the mineral resources that underpin mass-scale shifts to electric and renewable energy solutions. To take one example, world copper production needs to double in the next decade in order to meet the expected demand (Bonakdarpour & Bailey, 2022). This means more orebodies need to be found, as existing orebodies will be unable to provide this level of production.

Searching for new orebodies is, however, a notoriously difficult activity and a high-risk, high reward investment option. What are some of the key challenges that mineral explorers face?

Geological systems are highly non-linear, with processes that operate across scales, from molecules to cratons, and across a full range of psycho-chemical conditions (Ord et al., 2012). For this reason, the clustering of metals into economic deposits requires a highly specific set of circumstances. Even should those circumstances have arisen, the metal accumulation needs to be preserved in the uppermost levels of the crust at explorable and mineable depths (Occhipinti et al., 2016).

Furthermore many geological terranes that are exposed at the surface have already been scoured for signs of mineralisation, consequently exploration has been increasingly shifting to areas where the prospective rock packages are buried beneath post-mineralisation cover materials. Exploration undercover is a major challenge that requires geophysical and geochemical techniques to be able to 'see though' the cover to investigate the subsurface geology. Geophysical techniques themselves vary greatly in their ability to image the subsurface in 3D, since each relies on one physical property of rocks, such as magnetism, gravity, conductivity or seismic velocity. Challenges in the application of these various methods, whether fundamental, such as the fact that potential field data are vertically integrated, or logistical, such as difficulty of deployment or long wait times for data return, mean that application of these data to exploration also has its own challenges.

For all of these reasons, mineral exploration is generally information-poor, compared at least to other environments such as financial market analysis, or social media profiling. Activities and processes that improve the quality of exploration data in the most strategic locations are therefore critical pathways to improve exploration success (Caers et al., 2022). As a result, a methodology such as passive seismic ambient noise tomography (ANT) that is able to produce depth-constrained images of the subsurface in 3D and which can be integrated with drill hole-scale lithological information is highly valuable. ANT, characterised by its cost-effectiveness, low environmental impact, scalability, and depth penetration, is emerging as a pivotal tool in both regional and local-scale mineral exploration. Although other geophysical methods, such as magnetotelluric (MT), induced polarisation (IP), electromagnetic (EM), and active seismic reflection imaging all provide improved information as a function of depth over their potential field counterparts, ANT stands out in terms of it's ease of deployment, low-impact and low cost making it a very scalable and versatile 3D imaging method.

Geoscientists have taken to describing the formation of mineral deposits in terms of mineral systems, which encompass the range of factors from the source of metals, the processes that transport those metals through to those that trap and preserve those metals (Wyborn et al., 1994; Hronsky and Groves, 2008; McCuaig et al., 2010). With a clear framework for a given mineral system, the exploration search space can be refined, conceptually at least, to provide at first large regions and later kilometre-scale tenements in which exploration can be conducted. But the question then becomes, what techniques can we use to speed up the decision making process by providing levels of confidence of 'prospectivity' for a given commodity within a given volume? In this paper we seek to provide one solution to the many challenges of exploration targeting by harnessing the new wave of computing power encapsulated in artificial intelligence.

Artificial intelligence (AI) stands as a transformative force across diverse industries, demonstrating an unparalleled capacity to create, predict, and optimise with extraordinary precision and efficiency. In the realm of mineral exploration, AI's ability to process and interpret complex data sets opens new avenues for identifying mineral deposits with enhanced accuracy. Through its adaptability and continuous learning, AI has begun to redefine traditional exploration processes. While its application has been explored in isolated instances of the exploration sequence—from prospectivity mapping to resource estimation—its potential in a comprehensive end-to-end approach remains largely untapped. By integrating AI with ambient noise tomography (ANT), we can significantly refine our approach to mineral prospectivity analysis. This integration allows for the fine-tuning of AI models to specific geological contexts, enhancing their predictive power for mineral deposit localization. Such a novel application of AI in mineral exploration, from initial prospectivity mapping to detailed resource estimation, heralds a new era in the sector. It promises a more targeted, efficient, and informed exploration strategy, crucial for meeting the increasing demand for minerals essential for the global transition to a sustainable energy future.

We show the promise of this method by first establishing a regional-scale picture of mineral prospectivity, using copper as our example, based on large-scale regional datasets referenced against the location of existing deposits. This approach is similar to many previous regional prospectivity models (Kreuzer et al., 2015; Skirrow et al., 2019; Lawley et al., 2021). We then utilise the scale-invariance of geophysical data to apply the same model to prediction of mineralisation at the scale of a single ore deposit. Finally, we take this deposit-scale model and re-integrate it at the district to province scale in our training data in order to refine our ability to predict and target mineralisation. This provides us with an end-to-end, probabilistic tool with which to make improved decisions for improving the success rate of exploration applied to the search for commodities critical to the energy transition.

## Methods

In the realm of mineral exploration, understanding the intricate 3D processes that govern mineral deposit formation is crucial. Deep-seated geological processes play a pivotal role in the aggregation of minerals, making 3D data indispensable for accurate exploration. However, historically, acquiring comprehensive 3D geophysical data has been a challenging endeavour, primarily due to the high costs and significant environmental impacts associated with traditional methods.

Enter ambient noise tomography (ANT), a geophysical imaging method that provides the first scalable, low-cost, and low-impact technique capable of delivering 3D subsurface insights. Its scale-invariance is particularly noteworthy, as it allows for its application across various stages of exploration—from broad, regional-scale prospectivity analysis to more detailed, high-resolution examinations at the deposit scale.

The 3D models generated through ANT provide an ideal foundation for applying generative AI methods, addressing a long-standing gap in the availability of suitable 3D data for such advanced analytical techniques in mineral exploration. Our methodological vision entails leveraging this synergy by training a foundational model using national-scale precompetitive data, aiming to illuminate regions with high mineral prospectivity. This model serves as a strategic tool during the tenement selection phase, subsequently fine-tuned at the prospect scale using high-resolution ANT surveys to identify precise drill targets.

As new drilling information becomes available, the model undergoes further fine-tuning, enhancing its predictive accuracy. This iterative process of local fine-tuning, and potentially integration into the base model through federated learning, ensures continuous improvement, enriching the regional prospectivity model over time.

By adopting this approach, we aim to transform the landscape of end-to-end mineral exploration, enabling a more efficient, data-driven, and environmentally conscious pathway to uncovering the mineral resources essential for our future.

*Ambient seismic noise tomography (ANT)*

By cross-correlating continuous seismic noise recordings between station pairs, ANT reconstructs estimates of the surface wave component of the seismic Green's function. This approach repurposes the seismic noise captured at various stations, enabling each station to act as a virtual active source for detailed subsurface exploration (Shapiro and Campillo, 2004; Curtis et al., 2006).

Historically, ANT has been instrumental in mapping the Earth's upper crust on a crustal and regional scale, primarily utilising low-frequency seismic waves originating from oceanic interactions with coastal areas (Shapiro et al., 2005; Saygin and Kennett, 2010; Ritzwoller et al., 2011; Chen et al., 2023). However, recent technological advancements in instrumentation and computational capabilities have facilitated the application of ANT at finer scales (Hand, 2014). By tapping into high-frequency seismic noise generated by meteorological and anthropogenic activities, the method has been adapted for more localised and detailed geological investigations (Lin et al., 2013; Hollis et al., 2018; Ryberg et al., 2021; Li et al., 2023; Jones et al., 2024).

In the realm of mineral exploration, ANT is increasingly recognized for its efficacy, especially in imaging deep subsurface features and areas covered by barren overburden, where traditional methods may falter. Fleet Space Technologies has pioneered the development of the Geode, the world's first seismic node engineered explicitly for mineral exploration (Olivier et al., 2022). Equipped with a lower-frequency geophone and edge processing capabilities, the Geode minimises data transmission rates, enabling real-time data relay via low-power, long-range, direct-to-satellite IoT communications. Subsequent cloud processing delivers 3D subsurface models with unprecedented speed, allowing vast areas to be imaged with high resolution in a matter of days (Jones et al., 2024). In mineral exploration, where the rate of new discoveries is directly related to how fast we can traverse a large search space and make effective decisions, rapid and dynamic imaging of the subsurface is a potentially transformative development that could provide the missing ingredient required to make the discoveries required to transition to renewable energy.

Another consequence of the rapid nature of Fleet's ANT surveys is the massive amounts of 3D data we are amassing. With over 300 surveys completed in less than two years, this wealth of 3D data forms an ideal substrate for developing new generative AI methods for mineral exploration. Since ANT is scale invariant, we can also incorporate local high resolution surveys with regional-scale data, paving the way for a comprehensive, end-to-end mineral exploration methodology that leverages the full potential of modern seismic imaging and artificial intelligence.

*Artificial intelligence*

AI has caused a transformative shift across various industries, due to its ability to create, predict, and optimise with unprecedented precision and efficiency. The current era of artificial intelligence is characterised by its ability to generate new data that resembles the training data, and has revolutionised fields ranging from healthcare, where it aids in drug discovery and personalised medicine, to entertainment, where it crafts realistic digital content (Fui-Hoon et al., 2023). Central to its utility is the concept of fine-tuning, where a pre-trained AI model is adapted to specific tasks or datasets, enhancing its performance and applicability to niche domains (e.g. Too et al., 2019).

This process allows for the customisation of AI models to address particular industry challenges or opportunities, making the technology incredibly versatile and powerful. Furthermore, continuous learning enables these AI systems to evolve and improve over time, integrating new data and experiences to refine their predictions and outputs. This aspect of generative AI ensures that its applications remain cutting-edge, adapting to changing environments and accumulating knowledge, much like human learning.

In the context of mineral exploration, machine learning and generative AI has shown promise in recent years during discrete steps in the exploration timeline: in early stage exploration for prospectivity mapping (Rodriguez-Galiano et al., 2015; Albrecht et al., 2021; Woodhead & Landry, 2021), for tackling geophysical inverse problems during target delineation (Olivier and Smith, 2023) and at the end of the exploration process during resource estimation (Dumakor-Dupey & Arya, 2021). However, a unified end-to-end approach, where a foundation model is continuously improved and fine tuned on data from a specific region or deposit as more data is collected has not yet been developed.

By harnessing the high-resolution, 3D subsurface models generated through ANT, generative AI could significantly enhance the prediction and identification of mineral deposits. Fine-tuning allows these AI models to be specifically adapted to the unique geological signatures and features relevant to individual mineral deposits whilst still maintaining a general understanding of mineral systems from its diverse training data. Continuous federated learning will ensure that the models improve as they ingest more data from subsequent surveys or drilling results.

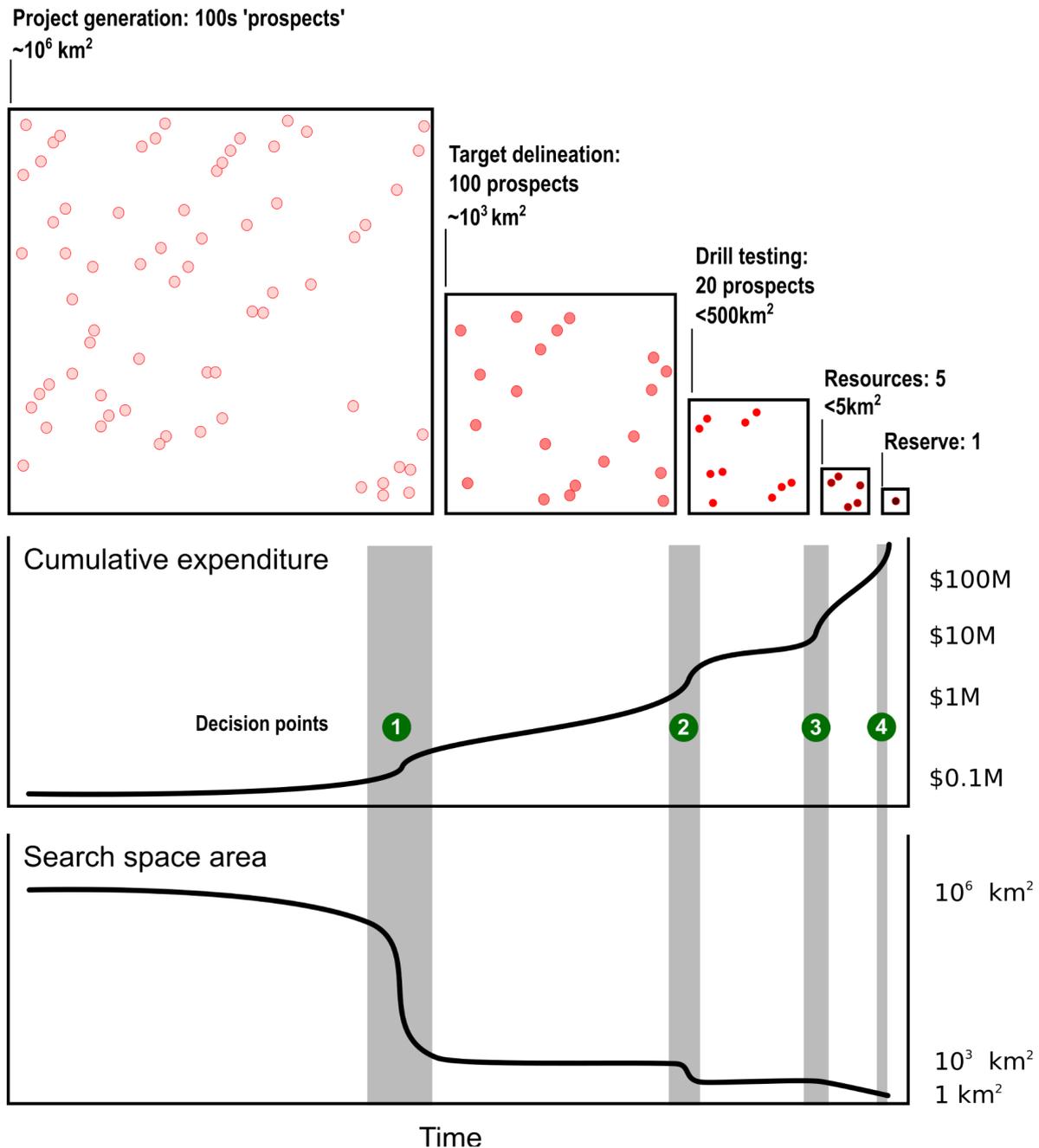

**Figure 1. The scale reduction process in mineral exploration towards mineral resource definition. Indicative cumulative expenditure increases along with concomitant decreases in search space area and key decision points are shown in lower panels. The exploration, discovery and delineation process for copper deposits take on average more than 12 years (Manalo, 2023), however in some cases it can take decades, for instance the discovery of Oak Dam West (King, 2019).**

This symbiosis between generative AI and ANT offers a new approach to mineral exploration, providing a powerful tool for identifying prospectivity and refining exploration targets with a level of detail and accuracy previously unattainable. The integration of these technologies could mark a new era in the search for critical minerals. In this study, we outline a proof-of-concept of this vision using a relatively simple end-to-end ML workflow coupled with pre-competitive national scale & proprietary local scale data.

## Results

*Continent-wide prospectivity model*

Our approach consists of building an Australia-wide base model to predict camp-scale copper prospectivity. The input geophysical data layers for this model are the Geoscience Australia compiled national magnetic, gravity and radiometric grids, along with 3D seismic velocity models from ambient noise tomography and body wave coda correlation. In addition to geophysical data, we also add the geographic coordinates (latitude, longitude) and the SRTM 30m DEM as features. Table 1 contains details of the datasets used. The gravity and magnetic data products include various pre-computed spatial filters that introduce a degree of spatial regularization into the dataset.

The 'true positive' training labels are all known economic primary copper deposits, as described in the S&P Capital IQ database. Defining cells that do not contain copper ('true negative') is more challenging, as we don't know definitively which cells do not contain copper. Our solution to this challenge is to consider all deposits in the S&P Capital IQ database that do not contain copper as either the primary or a secondary commodity as the 'true negative' labels. We argue that known deposits have undergone significant exploratory drilling, and not having reportable quantities of copper is a good indication of the lack of copper in such an area.

Given the challenge of defining accurate training labels in a given area, an effective machine learning strategy to tackle this problem is a tabular data approach based on features and labels of individual cells. Alternatively, one can use an image-based method, such as a convolutional neural network, based on a raster of tiles surrounding known deposits coupled with some rasterization strategy for the labels. For the illustrative purposes of this paper, we implemented the tabular approach using a gradient boosted trees algorithm (Chen & Guestrin 2016), with normalisation of data features and k-nearest-neighbours (KNN) imputation of missing data performed as preprocessing steps. Training of models in this class is computationally inexpensive for moderately sized datasets, and they scale well to problems with high input dimension (as is the case here, with 159 input features). However, tabular algorithms cannot take advantage of any spatial inductive bias that is not explicitly encoded within the feature set.

We trained the model using an 80/20 train/test split, with hyperparameters determined by 10-fold cross validation using an 80/20 train/validation split. To avoid merely interpolating the data geographically, we split both the train/test subsets and the hyperparameter cross-validation folds of the train subset using grouped sampling of the three character geohashes of the data. This process cuts out large contiguous geographic areas during both the cross-validation and test stages, strengthening the regularisation of fitting and promoting more geologically meaningful results.

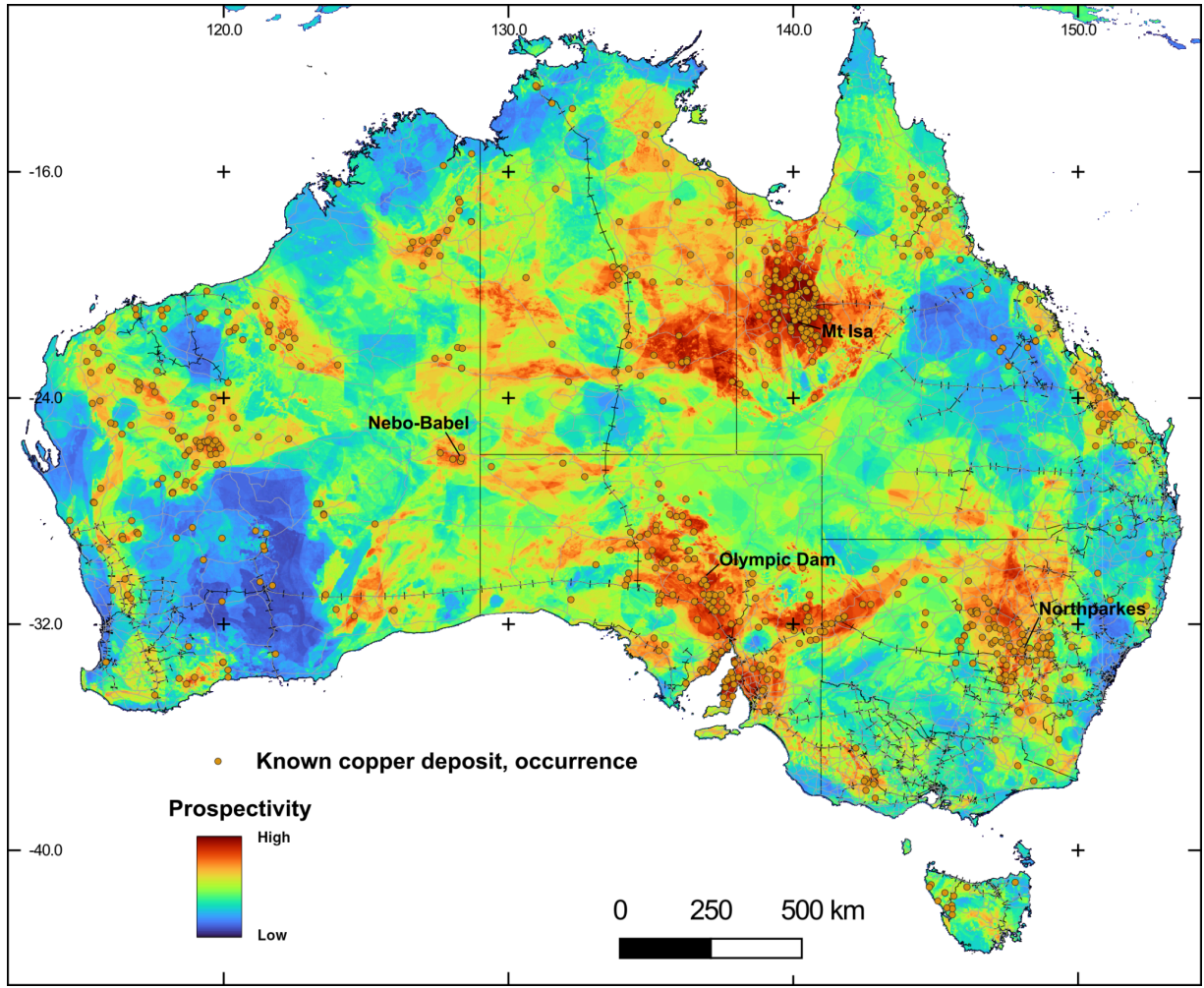

**Figure 2. Continental scale copper prospectivity prediction model, plotted with the known primary copper deposits as reported in the S&P Capital IQ database (2024). Note that the scale of prospectivity prediction is arbitrary due to measures taken to account for imbalanced classes during training.**

As a continent-wide prospectivity model, trained on a relatively limited number of input datasets, the consistent match between areas of high predicted prospectivity and known copper deposits and occurrences is encouraging (Fig. 2). While this correspondence is to some extent a natural outcome of the machine learning approach, it is further encouraging that zones of high prospectivity occur in regions of vastly different geological settings and despite coarse geographic partitioning of the training and test datasets. For example, while the model has predicted enhanced prospectivity in the vicinity of the Olympic Dam and Mt Isa deposits, which are themselves hydrothermal deposits within the broad iron oxide copper-gold (IOCG) deposit class, the model is also indicating high prospectivity in several regions around the West Musgraves region where magmatic nickel-copper sulphide deposits such as Nebo-Babel are located. Similarly, the porphyry copper-gold systems of the Macquarie Arc such as Northparkes are likewise indicated as high prospectivity.

As expected in a data-driven prospectivity analysis, the model is agnostic to geological processes. Petrophysical properties of dense, magnetic and sulphide-bearing rocks are similar enough across IOCG or magmatic nickel sulphide deposit types as to make the specific geological cause of those density or magnetic signatures and the absolute values of those signatures, redundant in this analysis. In addition, deep crustal architectural controls on ore deposit formation are also incorporated in this prospectivity modelling through the teleseismic and ANT data sets, which contribute the strongest overall controls on the prospectivity prediction (see Supplementary Fig.1). Location of major deposits adjacent to seismic discontinuities is a common theme in the Australian setting, and indeed globally (Begg et al., 2010; Hoggard et al., 2020; O'Donnell et al., 2023).

In terms of the broad tectonic elements of Australia, the model predicts that the North Australian and South Australian elements are the most strongly endowed regions of the continent in terms of copper prospectivity (Fig. 3a). These regions correspond to Proterozoic crust, much of which is likely underlain by modified Archean lower crust and mantle (Fraser et al., 2010; Hollis et al., 2010; Curtis and Thiel, 2019; Skirrow, 2022). Archean terranes of Western Australia are mapped as less prospective for copper than the Proterozoic-dominated terranes, while the Central Australian crustal element and Tasman crustal element have highly variable prospectivity (Fig. 3a).

In terms of the major geological terranes within these broader crustal elements, the Mount Isa Orogen, including the Cloncurry district, is a region that is highlighted as having very high prospectivity. The region is host to extensive alteration systems that encompass Mesoproterozoic IOCG and iron sulphide copper gold (ISGC) deposit styles (Williams, 1998; Williams and Pollard, 2001). The Warramunga Province has significant copper and gold-copper deposits in the Tennant Creek area (Skirrow and Walshe, 2002), and this is likewise highlighted in the prospectivity model. Interestingly, the Georgina Basin is identified as highly prospective, yet this region is perhaps the least well endowed with known mineralisation in the North Australian crustal element. Sediment-hosted replacement style base metal mineralisation in the Georgina Basin is predominantly Pb-Zn, however, copper systems may well be indicated as permissive in this analysis. Alternatively, it may be that the elevated prospectivity in this region is responding to the presence of copper deposits such as the Jervois deposit (McGloin et al., 2023), in the eastern Aileron Province (Arunta Region) as this province is in part overlain by the southern Georgina Basin.

Within the West Australian element, copper prospectivity is centred on the Paleoproterozoic Capricorn Orogen, which contains VMS-style base metal deposits including the DeGrussa and Red Bore (Pirajno et al., 2016; Agangi et al., 2018). Similarly, the Paterson Orogen, host to the Nifty, Telfer and Winu-Ngapakarra gold-copper and copper-gold deposits is also shown as a region of significant prospectivity. These deposits are replacement-style and intrusion-related, with possible causative magmatism being emplaced into host metasedimentary rocks during the Neoproterozoic (Anderson et al., 2001; Dalstra et al., 2023). Finally, we note the region of the western Yilgarn Craton also shows a corridor of enhanced prospectivity that trends north-south in a corridor that is near to the Jullimar Ni-Cu deposit hosted within Archean (2670 Ma) mafic-ultramafic intrusions (Lu et al., 2021). It is worth noting that Archean volcanic-massive sulphide deposits of the Yilgarn such as Gossan

Valley and Copper Bore (Hollis et al., 2015) are also highlighted, however the prospectivity in the central Yilgarn region is indicated as relatively limited beyond the known deposits.

The two major terranes of the South Australian element, the Gawler Craton and Curnamona Province have widespread prospectivity. Mesoproterozoic IOCG deposits of the eastern Gawler Craton including Olympic Dam (Ehrig et al., 2012) and replacement-style copper-gold deposits of the Curnamona Province such as Kalkaroo (Teale, 2006) are the main example deposit types in this region. Significantly, the model identifies copper prospectivity to be largely confined to the eastern Gawler Craton.

Much of the Central Australian crustal element has low to moderate prospectivity in this analysis. Exceptions are the regions around the central-western Musgraves where magmatic nickel-copper(-PGE) sulphide deposits are present, associated with the Warrakurna Large Igneous Province (Howard et al., 2015). Portions of the Albany Fraser Orogen is highlighted in the prospectivity model and is host to the mafic-ultramafic intrusions such as at the Ni-Cu-Co Nova-Bollinger deposit (Barnes et al., 2020) and similar rock packages across the Fraser Zone of the orogen (Spaggiari et al., 2015).

Finally, the Tasman element has three main regions of prospectivity highlighted. In the southwest, the Kanmantoo Province has prospectivity, extending from the Kanmantoo deposit itself (Oliver et al., 1998) into southwestern Victoria, and also the north-east to correlative rocks in New South Wales centred around the Koonenberry Belt (Greenfield et al., 2011). Prospectivity is also indicated in the Hodgkinson and Ertheridge provinces, north Queensland, and extending along the coastal hinterland of the   The main centre of prospectivity for the Tasman Element is focussed on the region around the porphyry deposits of the Macquarie Arc, such as Northparkes (Pacey et al., 2019).

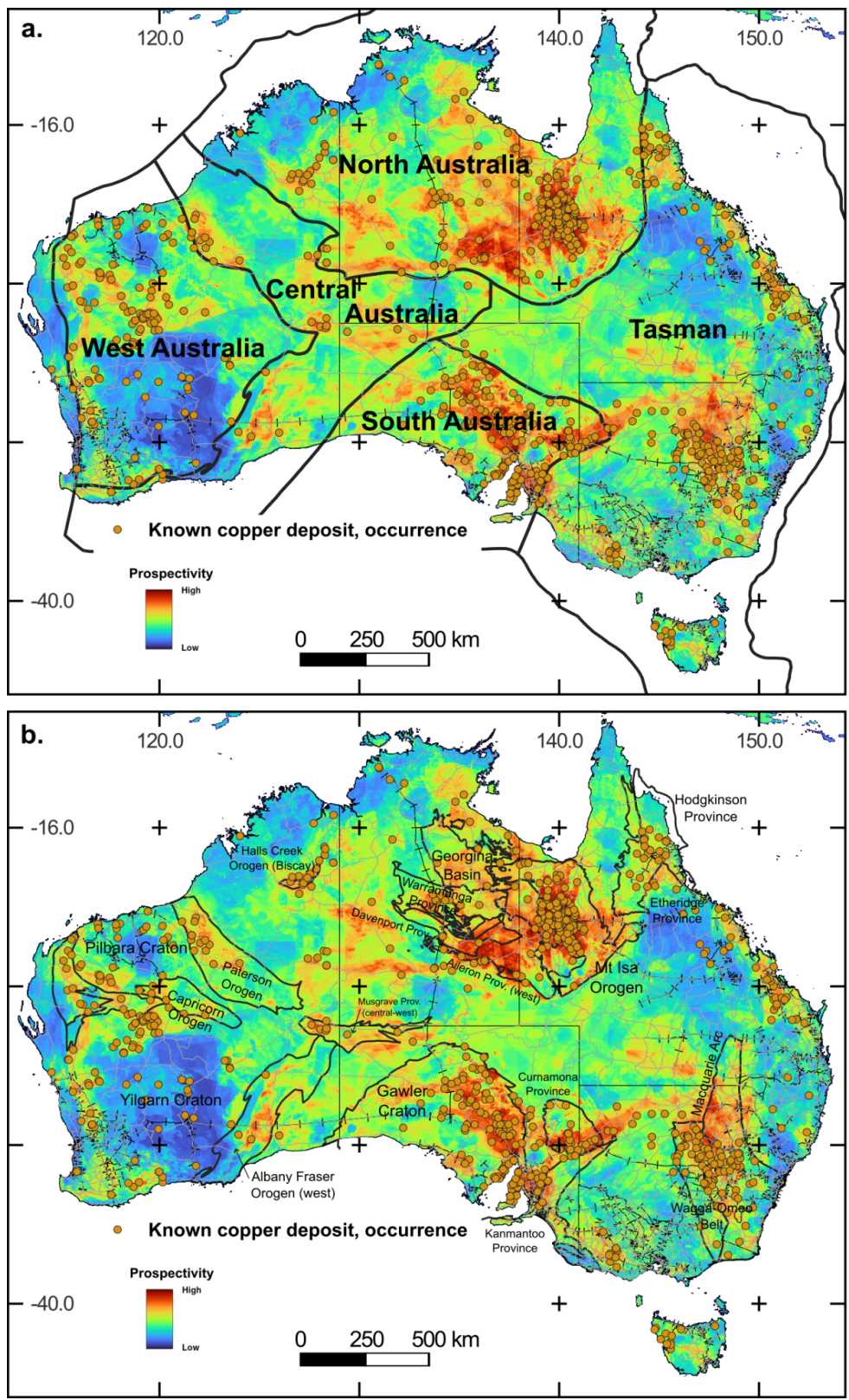

**Figure 3. Copper prospectivity model relative to the major crustal elements of Australia as mapped by Geoscience Australia (Shaw et al., 1996; Korsch and Doublier, 2016). a. Largest scale crustal elements of Australia. b. Selected terranes of Australia in areas of high model prospectivity.**

*District-scale observations from prospectivity model*

The model has been generated using an 80/20 geographical test/train split of the S&P Capital IQ mineral properties database. While the training dataset has its limitations, and some of the smaller prospects may be omitted that might otherwise be found in geological survey mineral occurrence databases, to a first order the ML prediction defines many areas with known copper deposits as having high prospectivity. We highlight four examples of the regional-scale prospectivity analysis to showcase how the different input datasets influence the texture of the prospectivity output.

In the Yilgarn Craton, several VMS-style deposits are known, however, overall the prospectivity for copper as a target commodity is relatively low. VMS systems are a product of the basin architecture during formation of the volcanic systems, with extensional faults and focussed fluid flow required to form deposits, and those deposits often having a strong strataform control, being localised at the stratigraphic group level (Hollis et al., 2015). The geophysical expression of such systems is however complicated by the overprinting events that have affected the craton, including cratonisation process itself. The orogenic events may have dissected stratigraphic units, meaning the magnetic signature of prospective units may be quite limited, at least to the resolution of a national-scale magnetic grid on which the algorithm way trained. Likewise, the geophysical expression of early fault systems that formed these deposits may have been impacted by broader-scale homogenisation of the lower and mid crust that has occurred during the late granite 'bloom' (Czarnota et al., 2010). These crustal-scale melting events have implications for the seismic velocity structure of the craton, which in turn means that larger-scale geophysical data such as the ANT may be indicating lower prospectivity given, for example, the higher wave speeds across a larger region.

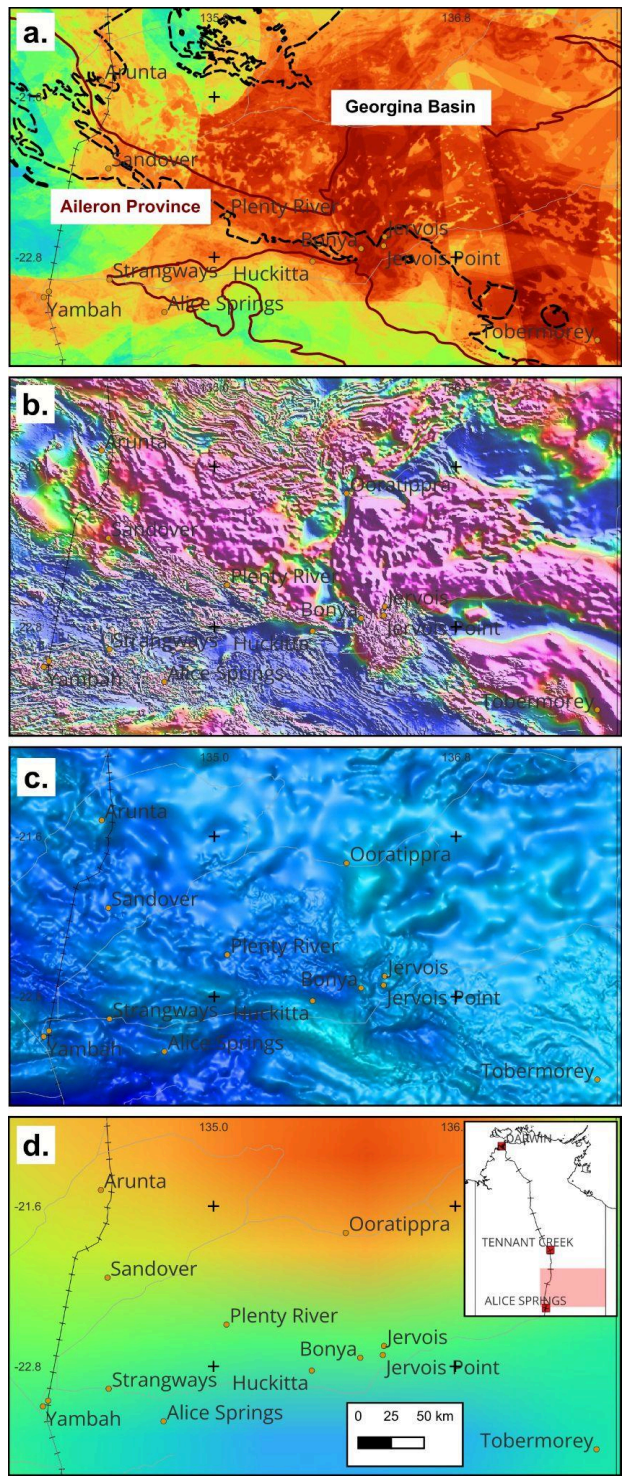

**Figure 4. Detail of the interface between the Paleoproterozoic Aileron Province and the overlying Neoproterozoic to Cambrian southern portion of the Georgina Basin and a comparison with different input datasets and the copper prospectivity model. Inset shows the location of the mapped area with respect to the Northern Territory. a. Prospectivity model with the outlines of the Aileron Province and Georgina Basin. b. National scale total magnetic intensity image, Geoscience Australia. c. National scale Bouger gravity image, Geoscience Australia. d. Example of one of many seismic data inputs to the prospectivity model, the ANT Vs model at 70km (Chen et al., 2021).**

In the case of the southern Georgina Basin and underlying Aileron Province, a large region appears prospective in the model, and this region is comparatively devoid of known deposits (Fig. 4). The training data includes the Jervois and Bonya deposits, which are stratabound copper deposits hosted within calc-silicate rocks with mineralogy including magnetite, garnet and pyroxene (McGloin et al., 2023). Such deposits train the algorithm to see dense, magnetic rocks as prospective. The continental-scale magnetic intensity data are however relatively low resolution to the immediate north of the northern Aileron Province around the Jervois and Bonya deposits, in part because the basement becomes deeper in this region due to burial by the southern Georgina Basin. Several gravity high 'ridges' extend in the region north of the deposits which likewise suggest prospectivity to the model. In terms of the seismic data inputs, there are many layers at many depths that are incorporated into the model. Nevertheless, for illustration we highlight the ANT velocity model at a depth of 70km, which has a relatively steep gradient in this region from a high Vs zone to the north to a lower Vs zone in the south, with this gradient zone being where the known copper deposits are located.

The texture of the prospectivity model in the Macquarie Arc in the Lachlan Orogen (Fig. 5) shows somewhat greater granularity than the Aileron Province/Georgina Basin example. This region is host to porphyry-style copper and copper-gold-molybdenum deposits, which are typically associated with magnetic anomalies caused by magnetite-bearing granites. The gravity signal is similarly textured in this region, related to the relatively shallow basement and complex relationships between stratigraphy and intrusive rocks. In addition, in this region, shallow datasets such as radiometrics may also be influencing the prospectivity significantly, with areas of outcrop that are devoid of copper deposits in the training data appearing as relatively low prospectivity zones. Finally, it is worth noting that although the Cadia hydrothermal, intrusion-related gold-copper deposit (Holliday et al., 2002) was not used as a training point, the region around Cadia ranks as relatively high prospectivity for copper in the prospectivity model. In this region, it appears the host volcanic rocks and monzonite-diorite porphyry stocks of Cadia are geophysically similar to porphyry copper-gold deposits such as Northparkes, so as to be highlighted as prospective in this model.

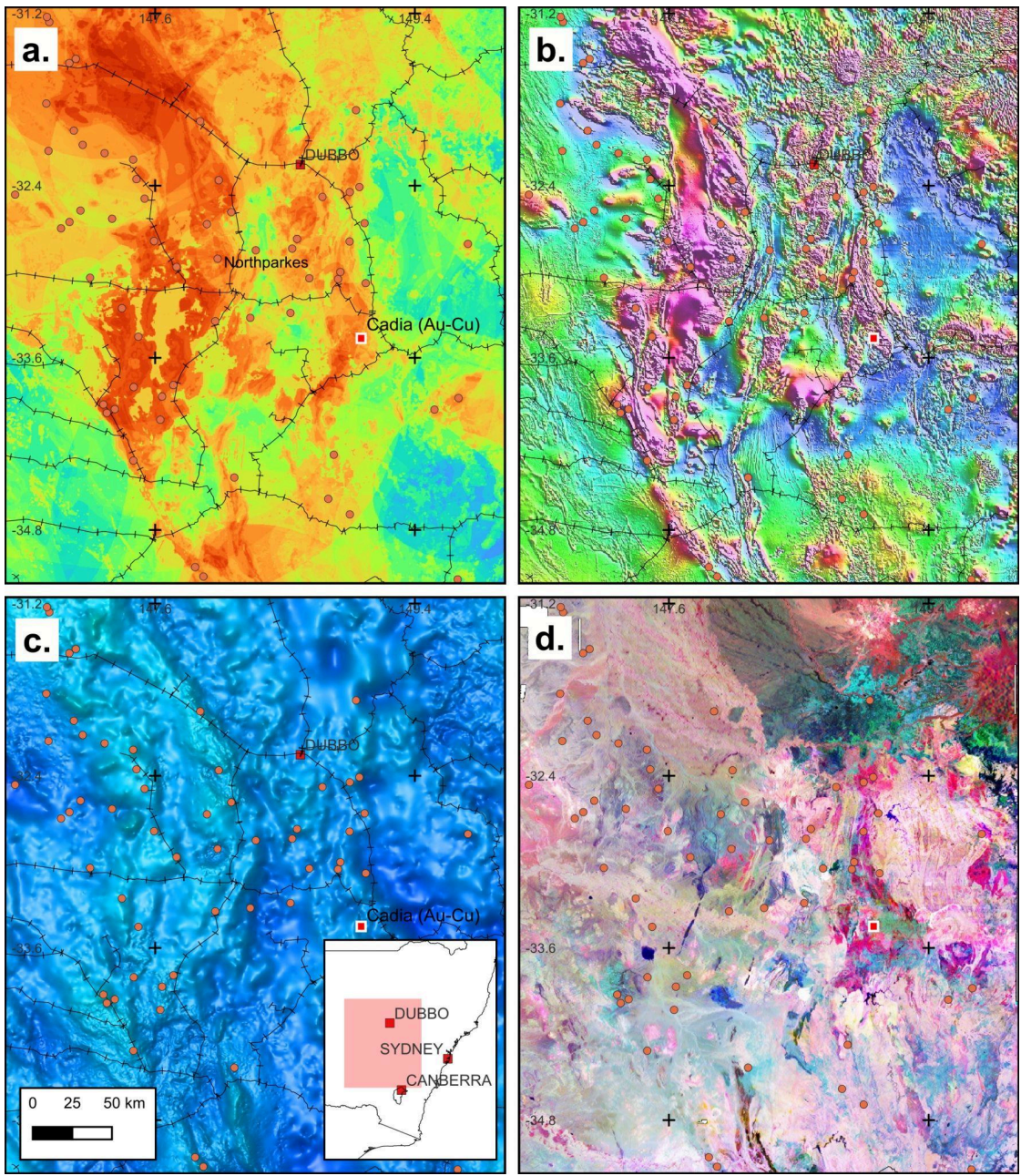

**Figure 5. Detail of prospectivity model and training datasets in the vicinity of Dubbo. Inset shows location. Cadia deposit is shown for reference, but was not utilised as part of the training dataset. a. Prospectivity model. b. National scale total magnetic intensity image, Geoscience Australia. c. National scale Bouger gravity image, Geoscience Australia. d. Ternary radiometrics image, Geoscience Australia.**

Finally, we note that the area of the Yorke Peninsula, in the southeastern Gawler Craton appears as one of the highest zones of prospectivity in the model (Fig. 6). This high prospectivity may relate to the overall strong magnetic response of the metasedimentary rocks (Wallaroo Group) and Mesoproterozoic intrusions that host copper-gold mineralisation in this region (Conor et al., 2010). The gravity signature of the region is similarly broad, although the north-western area of Yorke Peninsula has more texture than the central region.

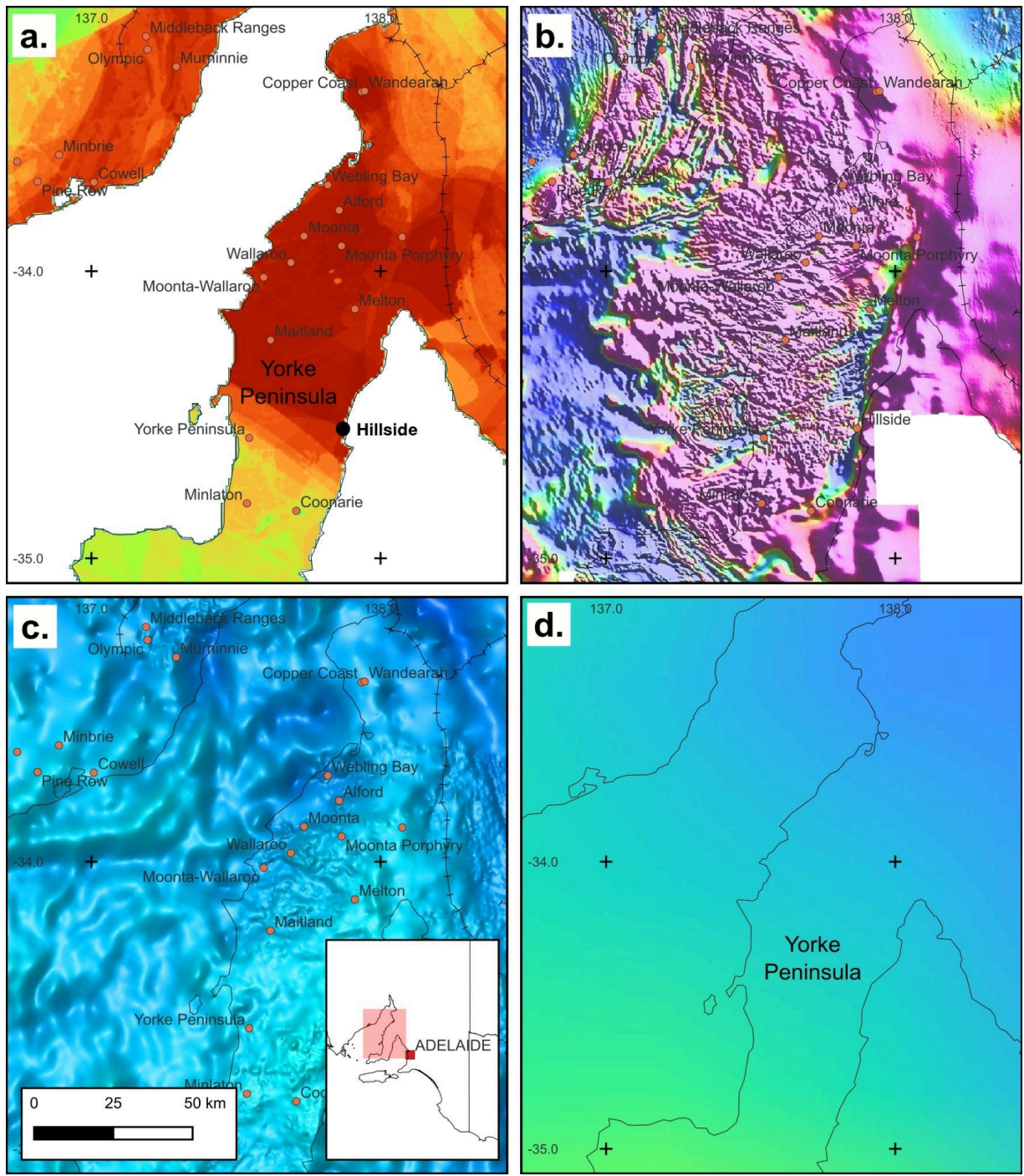

**Figure 6. Detail of prospectivity model and training datasets in the vicinity of the Yorke Peninsula, South Australia. Inset shows location. a. Prospectivity model. b. National scale total magnetic intensity image, Geoscience Australia. c. National scale Bouger gravity image, Geoscience Australia. d. Example of one of many seismic data inputs to the prospectivity model, the ANT Vs model at 30km (Chen et al., 2021).**

In addition, seismic data in this region is dominated by long wavelength features having little texture. Each of these elements mean that the overall prospectivity is 'high', however, this may reflect the nature of the available input data more than the true prospectivity of the region. This is an excellent example of a region where further refinement of the prospectivity model using more localised, high-resolution and site-specific datasets would be expected to improve the model resolution.

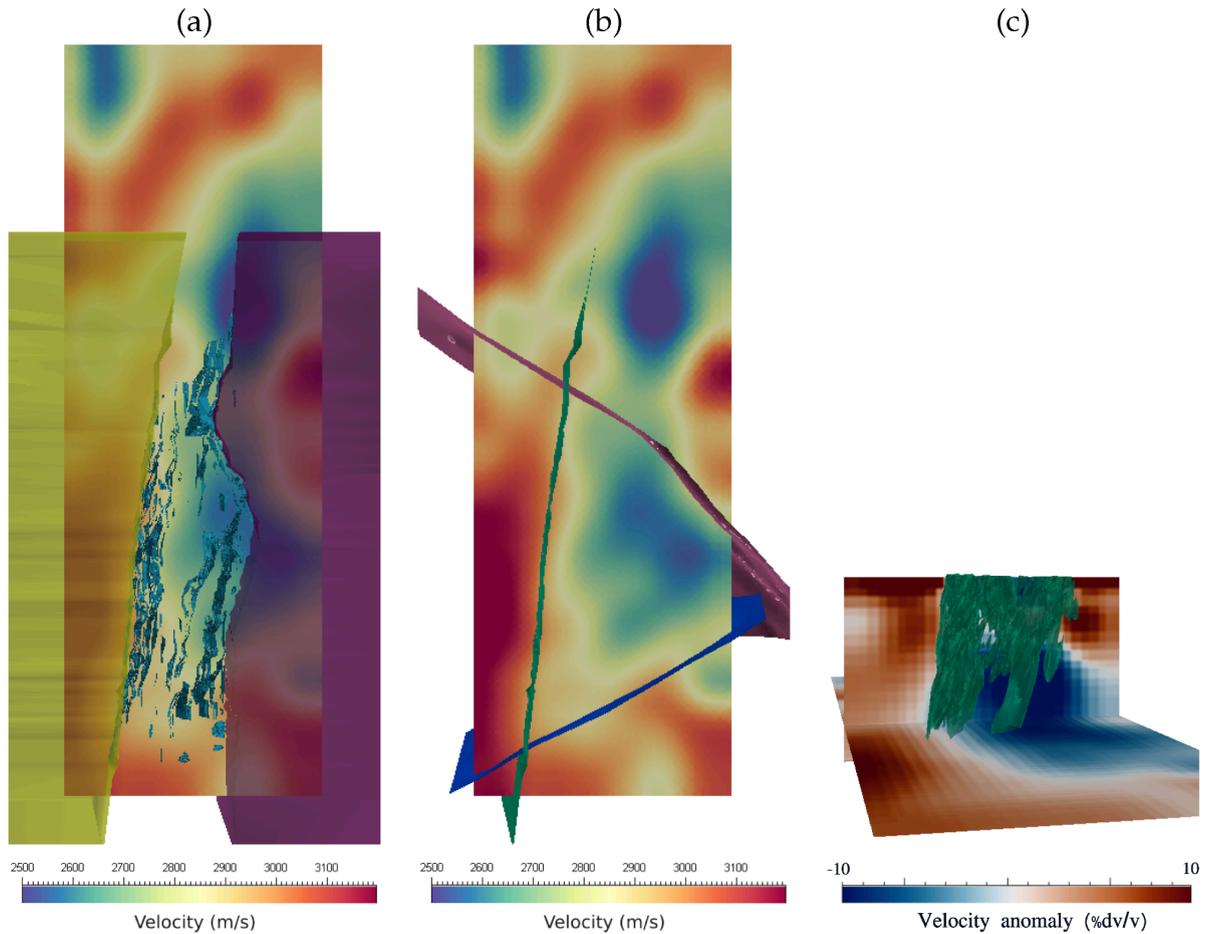

**Figure 7. Slice through ANT velocity model at −400 mRL (sea-level) shown against the hanging wall, footwall, and 0.20% copper mineralisation shells taken from the geological model for Hillside, Rex Minerals. (b) Slice from (a) shown against major faults mapped by Rex minerals interpreted to have a control on mineralisation. (c) Cross-sectional view of ANT velocity anomaly model (defined in text) along latitude 6175030 (MGA Zone 53 (GDA94)), showing a package of modelled skarn occupying the moderate-to-low velocity domain that defines the PPSC. Figure and caption from (Jones et al., 2024).**

## *Local fine-tuning: Hillside IOCG deposit*

Given the scale invariance of geological features in mineral systems, along with the scale invariance of ANT, one can fine tune the base model by training on local data. This approach will preserve the learned geological and geophysical features and knowledge from the base model to improve performance of the predictions on the local prospectivity model. The power of this approach is that we are able to construct a local prospectivity model from sparse local data by leveraging the learnings from the base model. We investigate this approach by considering the Hillside IOCG deposit, situated on the Yorke peninsula which was one of the highest zones of prospectivity in the continent-scale model. This deposit was also the location of a recent ANT survey (Jones et al., 2024), making it the ideal location for demonstrating the potential of fine-tuning the base model.

The Hillside copper-gold deposit is an example of a transitional magnetite-hematite style iron oxide copper-gold (IOCG) deposit (Conor et al., 2010). The deposit contains a published ore reserve estimate of 186Mt @ 0.53% Cu and 0.14g/t Au, containing 989kt of copper and 834koz of gold (Rex Minerals, 2022). Copper sulphide is dominated by chalcopyrite with subordinate bornite and chalcocite. Alteration mineralogy is dominated by early K-feldspar, garnet, pyroxene and magnetite-dominated assemblages that are variably overprinted by sulphide-bearing amphibole, chlorite, white mica and hematite-dominated assemblages, that formed within a steeply dipping shear zone and fault network (Conor et al., 2010; Ismail et al., 2014). Gabbroic and granitic intrusions were also part of the thermal driver for mineralisation and are themselves variably overprinted by the alteration minerals.

Hillside deposit sits within the Olympic Cu-Au Province, a metallogenic province along the eastern margin of the Gawler Craton (Skirrow et al., 2007; Reid, 2019). The Olympic Cu-Au Province links Cu-Au(+U) deposits that span a spectrum of physical and chemical deposit types, with the largest deposits being hematite breccia systems such as Olympic Dam, Prominent Hill and Carrapateena (Ehrig et al., 2012). The spectrum of IOCG deposit types across the eastern Gawler Craton, and indeed across other IOCG terranes (Corriveau et al., 2016), mean the exploration search space for IOCG deposits is complex and from a geophysical perspective encompasses both potential field methods looking for either magnetic highs, gravity highs, or more typically some combination of both. Seismic velocity information from the Hillside deposit suggests copper mineralisation occupies regions of moderate shear wave velocity (2600 - 2900 m/s) that appears to sit in a transitional zone between high and low velocity domains (Jones et al., 2024).

To construct a database of true positive labels, we use a section of the resource estimate model for the Hillside deposit. Similar to the base model, we consider each cell in the model that intersects the deposit as prospective. The difference here is that the prospective cells here represent small segments of the total deposits, whereas the prospective cells in the base model correspond to entire deposits. As more information is gathered about the deposit the model is further fine tuned. In the process the model transitions from drill targeting and optimisation, to a resource estimation model. By following these steps, details from the national model are retained, allowing for fine-tuning and adaptation to local datasets while benefiting from the strengths of both approaches. Reference predictions using the national model only are shown in Supplementary Fig. 2, and show some agreement with the distribution of mineralization, however it is clear that local training is required to reach the level of spatial accuracy needed for drill campaign planning.

To continue our proof-of-concept hierarchical modelling to the local scale, we specifically investigate the potential for combining regional and local features to predict the surface projections of the 0.2% Cu ore shells (as supplied by Rex Minerals). The local feature set comprises the leaf-index encoded national-scale data as categorical features (He et al., 2014), local gravity and total magnetic intensity grids (https://map.sarig.sa.gov.au/), and an ANT derived seismic shear-wave velocity model spanning the range from 1053 m below sea level to the surface.

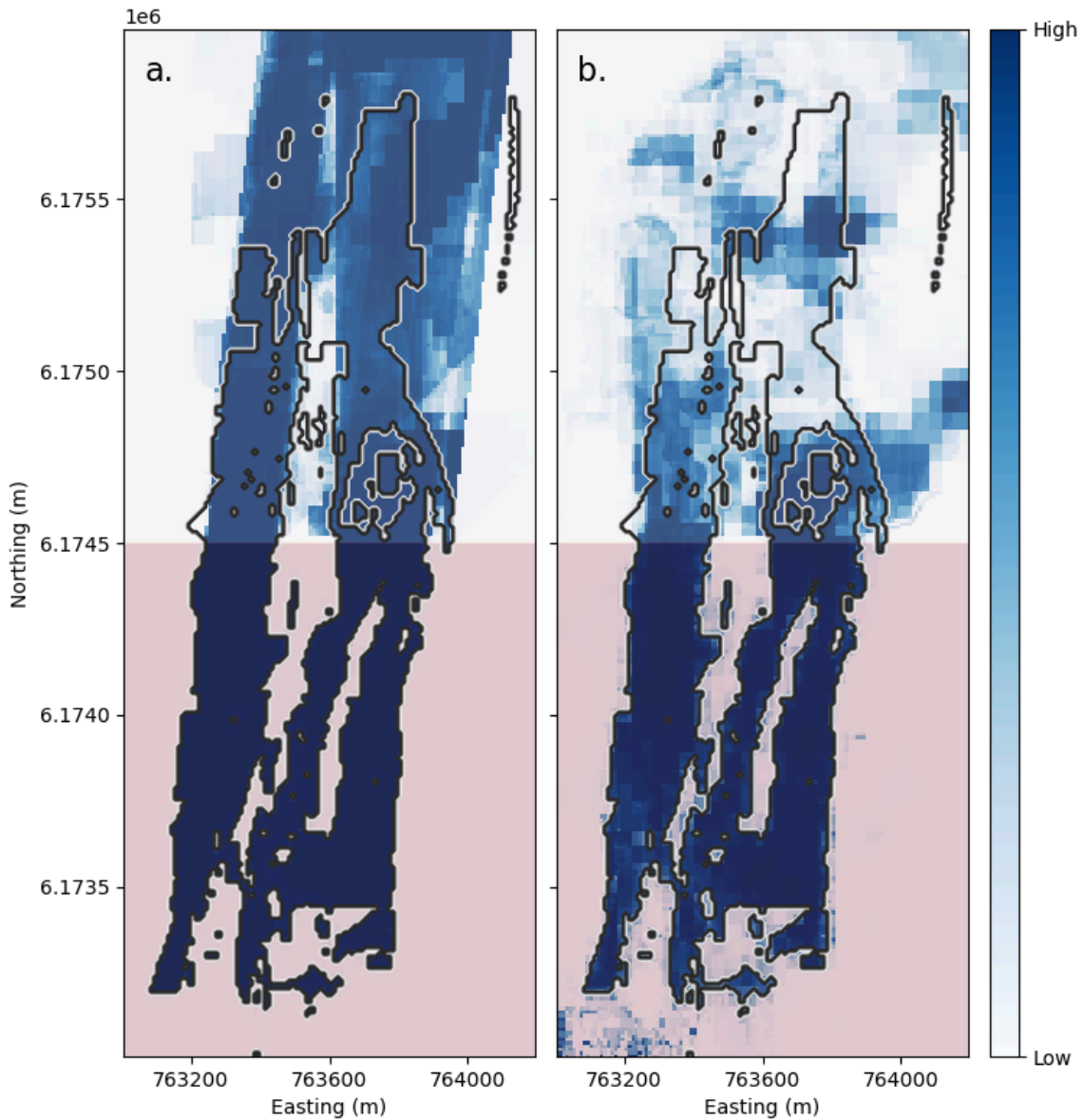

**Figure 8. Locally trained ore shell prediction models for the "prospect extension" setting, with a. showing results including both geophysical and geometric features, and b. geophysical features only. Ore shells are shown by grey contours, and the region used for training data is shown by red shading (the southern 50% of the study area).**

Using the same workflow as the national-scale prospectivity grid (normalisation of non-categorical features, KNN imputation of missing data, gradient boosted tree classification), we investigated two data settings within the same geological case study: a "prospect extension" setting, where we assume detailed knowledge of a part of the study area and wish to predict further mineralization proximate to the well known part; and a "greenfield" setting, where we assume sparse scattered measurements and wish to provide the best mineralization estimate between them to facilitate efficient planning of a delineating drill campaign. In the first case, we assume knowledge of the target within the southern 50% of the study area as training data; in the second case, we divide the study area into 100 x 100 m squares and use 20% as training data. In both cases, we then employ 10-fold grouped cross-validation using a 25/75 training/validation split for hyperparameter tuning. We further investigated adding geometric features to the data to promote continuity of the recovered ore shells.

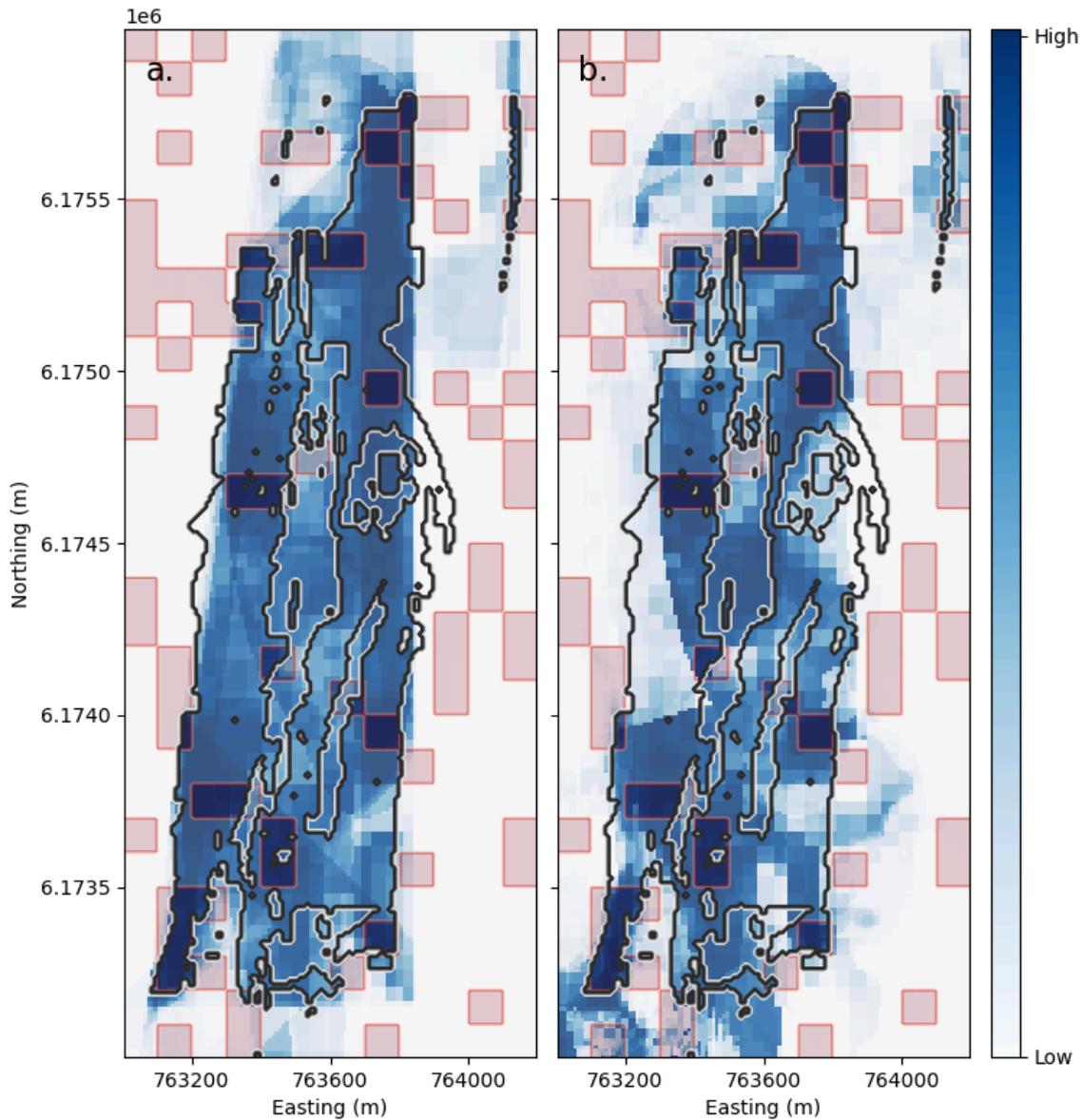

**Figure 9. Locally trained ore shell prediction models for the "greenfield" setting, with a. showing results including both geophysical and geometric features, and b. geophysical features only. Ore shells are shown by grey contours, and the region used for training data is shown by red shading (random 100 x 100m patches comprising 20% of the study area).**

Figure 8 shows the results for the prospect extension setting, with 8a. showing the prediction results for all features (including geometric ones) and 8b. showing predictions using geophysical features only. The predictions have F1 scores of 0.60 and 0.55 respectively for this case. We can see that the inclusion of geometrical features allows for predictions that capture the general N/S trend of the mineralization well, but result in spurious predictions at the extreme north of the study area where no training data is available to terminate the prospective region. Results from geophysical features only generally decay in accuracy away from the training data but also conform to the mineralization region.

Figure 9 similarly shows the results for the greenfield exploration setting. As expected, the performance of predictions is better in this setting (F1 scores of 0.77 and 0.68 respectively), due to more even distribution of training data, despite the overall volume of training data being only 40% of that in the prospect extension setting. The limitations associated with explicit geometrical features are apparent from Fig. 9a. where spurious sharp edges are generated even in the interpolation context. Further work could investigate image-based incremental learning strategies to incorporate hierarchical context within a more powerful model class than decision tree based models (e.g. Roy et al. 2020).

Both cases highlight the potential of hierarchical modelling, combining the continental scale background information with local geophysical features to take preliminary or area-limited drill results and make predictions for the distribution of unseen mineralization. Utilisation of this information to improve the efficiency of a drill-planning campaign promises to substantially reduce the overall cost of exploration. Given the relatively limited availability of training labels for multiscale mineral prospectivity modelling (known deposits and their geometries) compared to the large quantities of available unlabeled geophysical data, substantial performance improvements to the AI modelling framework are in particular likely to arise from self-supervised or contrastive pre-training (Radford et al. 2021, Ravula et al. 2021) of multivariate geophysical datasets before fine-tuning on prospectivity prediction tasks.

**Conclusions**

The novel integration of ambient noise tomography (ANT) and artificial intelligence (AI) demonstrated in this study represents a significant advancement in the field of mineral exploration. By employing this integrated approach, we have successfully enhanced the precision and efficiency of identifying and delineating mineral resources, which are pivotal for the transition to a sustainable, low-carbon economy.

Our results underscore the power of combining advanced geophysical imaging with modern AI. We show a new continent-wide data-driven prospectivity model for copper that aligns with known copper deposits and showcases the potential in various geological settings across Australia.

We then show that this model can be fine-tuned by focusing on the Hillside IOCG deposit, where the use of localised high-resolution ANT data to fine-tune the continent-wide prospectivity model substantially improved the accuracy of orebody delineation. This end-to-end methodology not only expedites the exploration process but also fosters a more sustainable and informed approach to mineral resource development.

The success of this integrated method paves the way for future exploration endeavours. It suggests a scalable model that could be adapted to various mineral types and geological contexts, enhancing our ability to meet the growing demand for essential minerals. Continuous improvement and expansion of our AI models, incorporating evolving geological and geophysical data, will remain a focal point, ensuring that our exploration strategies remain at the forefront of technological advancement. This study not only contributes to the field of geosciences but also offers a tangible pathway toward the responsible and efficient utilisation of our planet's mineral resources.

## Acknowledgements

The authors would like to thank Rex Minerals for providing access to proprietary data in this study.

## Data Availability

Data for the national prospectivity model is available from the permanent identifiers associated with their references, detailed below in Table 1. The geophysical data used for the Hillside case study is available at SARIG (https://map.sarig.sa.gov.au/).

| Dataset | Layers | Resolution | Source |
|---|---|---|---|
| GA Australian National Gravity Grid 2019 | 10 | 400m | Lane et al. 2020 |
| GA Australian National Magnetics Grid 2019 & enhanced products | 22 | 80m | Poudjom Djomani et al. 2020<br><br>Morse 2020 |
| GA Australian National Radiometrics Grid 2019 | 12 | 100m | Poudjom Djomani et al. 2020 |
| AuSeis teleseismic coda autocorrelation model | 80 | 0.5° | Qashqai et al. 2019 |
| CSIRO Ambient Noise Cross-Correlation model of Australia | 31 | 0.75° | Chen et al. 2021 |
| GA Shuttle Radar Topography Mission digital elevation model of Australia | 2 | 30m | Gallant et al. 2011 |

**Table 1: Datasets used in the generation of the national-scale prospectivity model (additionally using latitude and longitude as geometric constraints).**

**Supplementary Figures**

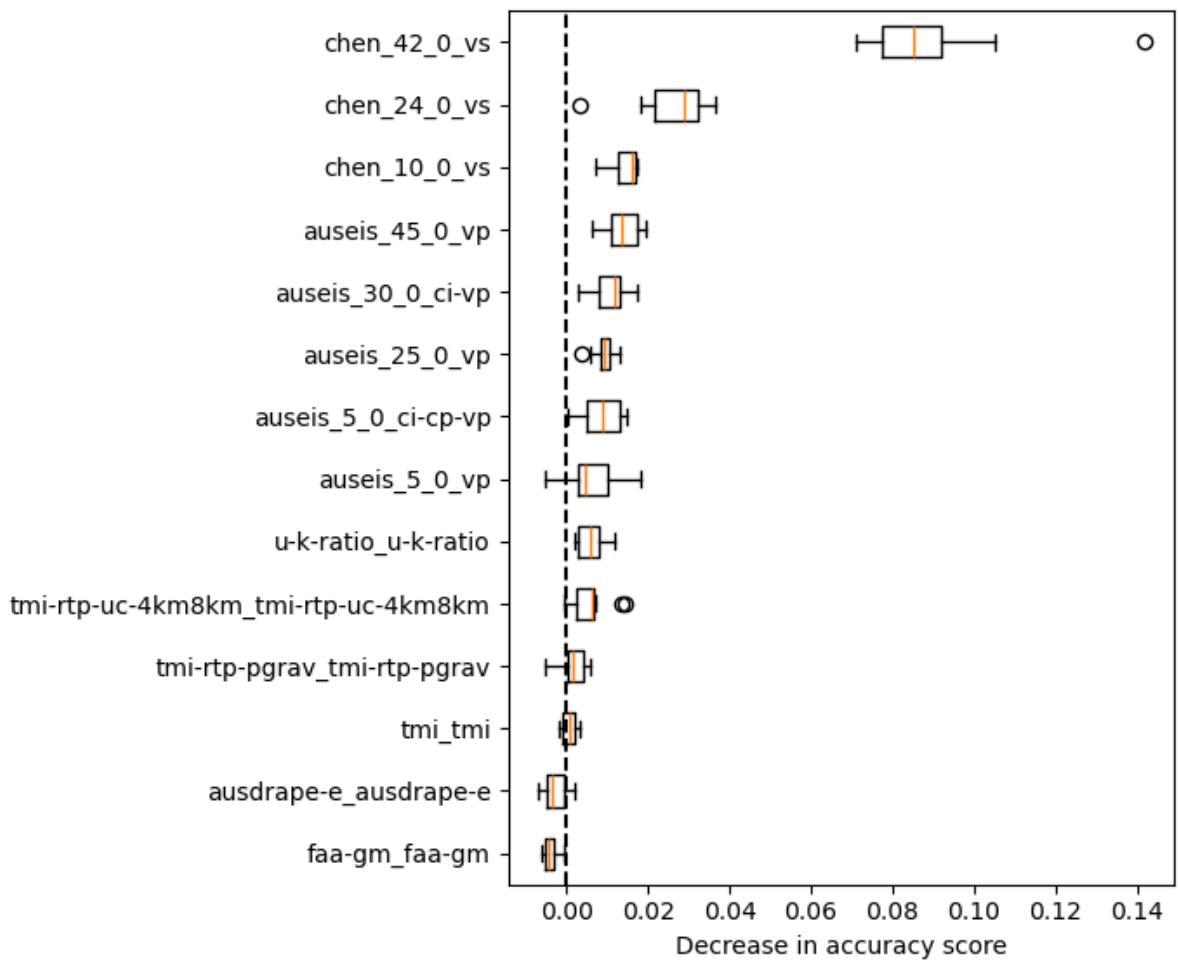

**Supplementary Figure 1: Permutation importances of the image patch central values of representative features chosen by hierarchical clustering analysis. The chen_ and auseis_ features represent large-scale crustal structure imaged using ambient-noise cross-correlation tomography (ANT) and teleseismic coda correlations respectively, and have the largest overall impact on mineral prospectivity at continental scale.**

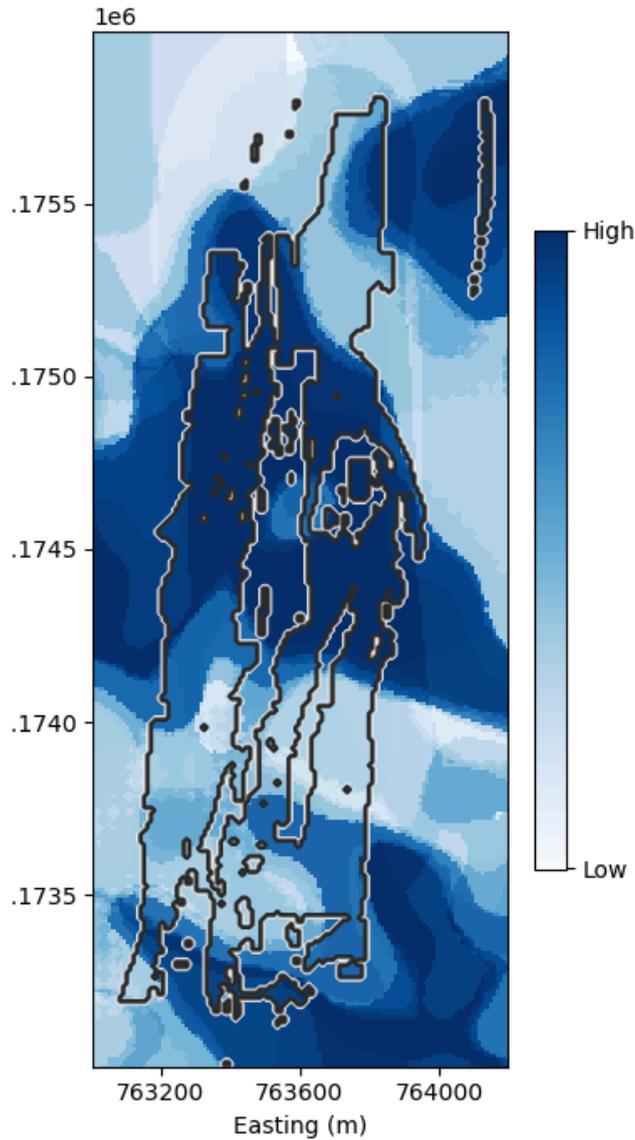

**Supplementary Figure 2: Reference "untrained" prospectivity prediction obtained by applying the national prospectivity model directly to local geophysical data at Hillside. To obtain relevant seismic velocity scalings, the local ANT model is stretched to the mean Australian crustal thickness (40km) and applied as perturbations to the Chen 2024 model.**